\documentclass{article}

\usepackage{arxiv}

\usepackage[utf8]{inputenc} 
\usepackage[T1]{fontenc}    
\usepackage{hyperref}       
\usepackage{url}            
\usepackage{booktabs}       
\usepackage{amsfonts}       
\usepackage{nicefrac}       
\usepackage{microtype}      
\usepackage{lipsum}
\usepackage{amsmath,graphicx,times,float}

\title{POLYPHONIC SOUND EVENT AND SOUND ACTIVITY DETECTION:\\ A MULTI-TASK APPROACH}

\author{
  Arjun Pankajakshan\thanks{AP is supported by a QMUL Principal's studentship. HLB is funded under EPSRC grant EP/R01891X/1. EB is supported by RAEng Research Fellowship RF/128 and a Turing Fellowship. This research was supported by an NVIDIA GPU Grant.} \\
  School of EECS\\
  Queen Mary University of London, UK\\
  \texttt{a.pankajakshan@qmul.ac.uk} \\
   \And
  Helen L. Bear \\
  School of EECS\\
  Queen Mary University of London, UK\\
  \texttt{h.bear@qmul.ac.uk} \\
  \And
  Emmanouil Benetos \\
  School of EECS\\
  Queen Mary University of London, UK\\
  \texttt{emmanouil.benetos@qmul.ac.uk} \\
}

\begin{document}
\maketitle

\begin{abstract}
Polyphonic Sound Event Detection (SED) in real-world recordings is a challenging task because of the dynamic polyphony level, intensity, and duration of sound events. Current polyphonic SED systems fail to model the temporal structure of sound events explicitly and instead attempt to look at which sound events are present at each audio frame. Consequently, the event-wise detection performance is much lower than the segment-wise detection performance. In this work, we propose a joint model approach to improve the temporal localization of sound events using a multi-task learning setup. The first task predicts which sound events are present at each time frame; we call this branch `Sound Event Detection (SED) model', while the second task predicts if a sound event is present or not at each frame; we call this branch `Sound Activity Detection (SAD) model'. We verify the proposed joint model by comparing it with a separate implementation of both tasks aggregated together from individual task predictions. Our experiments on the URBAN-SED dataset show that the proposed joint model can alleviate False Positive (FP) and False Negative (FN) errors and improve both the segment-wise and the event-wise metrics.
\end{abstract}


\section{Introduction}
Sound event detection (SED) \cite{virtanen2018computational} is the task of detecting the label, onset, and offset of sound events in audio streams. The overlapping nature of different sounds interfered with noise makes it difficult for accurate detection of sound events. Broadly speaking, the term \textit{sound event} refers to a specific sound produced by a distinct physical sound source, such as `a car passing by', `a bird singing', `a gunshot' or `the melody of rain'. Different sound events have different duration and that can be dynamic too. For example, the event `rain' can be considered as a continuous event at the same time `a gunshot' is an instantaneous event. Typically, a soundscape contains multiple sound events that can occur simultaneously as in a real scenario or at separate time instants which are respectively termed as polyphonic sounds and monophonic sounds. 
In recent years, SED has been utilized in many applications including audio surveillance \cite{foggia2015reliable}, health care monitoring \cite{goetze2012acoustic}, urban sound analytics \cite{salamon2015feature}, bio-acoustics \cite{stowell2015acoustic, morfi2018deep} and smart home devices \cite{krstulovic2018audio}.

\subsection{Related Work}
\label{ssec:subhead}


Most recent advances in polyphonic SED are largely attributed to the use of Machine Learning and Deep Neural Networks \cite{kong2018audio,komatsu2016acoustic,zhang2012semi,benetos2017polyphonic,jansen2018unsupervised,kumar2017audio}. In particular, the use of Convolutional Recurrent Neural Networks (CRNNs) has significantly improved SED performance in the past few years \cite{cakir2017convolutional, xu2018large,ccakir2018end, adavanne2017sound}. However, there are three main disadvantages with current CRNN-based polyphonic SED approaches. 1) Most CRNN-based polyphonic SED systems use a frame-wise cost function for training; as a result these systems often show relatively high segment-wise accuracy but low event-wise accuracy. For example, in the DCASE 2016 task on event detection in real life audio \cite{dcase2016}, the F1 score is around 30\% at segment level (frame length of 1 second), but only around 5\% at event level (tolerance of 200 ms for onset and 200 ms or half length for offset). 
2) These methods require pre-segmented training data. For RNNs to make a prediction at every frame, it is necessary to provide the exact start and end times of the sound events in the training data, making data annotation an extremely time-consuming process. 3) CRNN-based SED requires post-processing methods to transform the output into event sequences. Within these limitations of CRNN based polyphonic SED, in this work we attempt to enhance CRNN based polyphonic SED performance using auxiliary learning without any additional input data representation using a joint model approach.

\subsection{Contributions of this work}
\label{ssec:subhead}

To the best of our knowledge this is the first attempt to treat the polyphonic SED task in a joint model framework with sound activity detection (SAD) as an auxiliary task. We define SAD as the task of detecting the presence or absence of any sound events, analogous to voice activity detection in speech processing. Our experimental results show that re-weighting the predictions of the SED model with the SAD model predictions, lessen False Positive (FP) errors in both segments and events, and False Negative (FN) errors in events which in turn improve polyphonic sound event detection performance at both the segment and the event levels. Implicitly, the joint model helps in better temporal localization of sound events. To claim the effectiveness of the proposed auxiliary task on polyphonic SED, irrespective of the benefits of multi-task learning, we also verify the proposed approach by training the SED and SAD models separately and re-weight the predictions. The results are compared with our baseline SED model, aggregation of separately trained SED and SAD models, and with the joint model.

The rest of the paper is organized as follows. Section 2 covers the proposed joint model with details on model architecture, feature extraction, training, and experiments. Section 3 presents the dataset and evaluation metrics used in this work. In Section 4 the results, and discussions over the results are reported. Section 5 provides conclusions and future directions related to this work.

\section{PROPOSED METHOD}
\label{sec:format}

\begin{figure}[ht]
  \centering
  \includegraphics[scale=0.6]{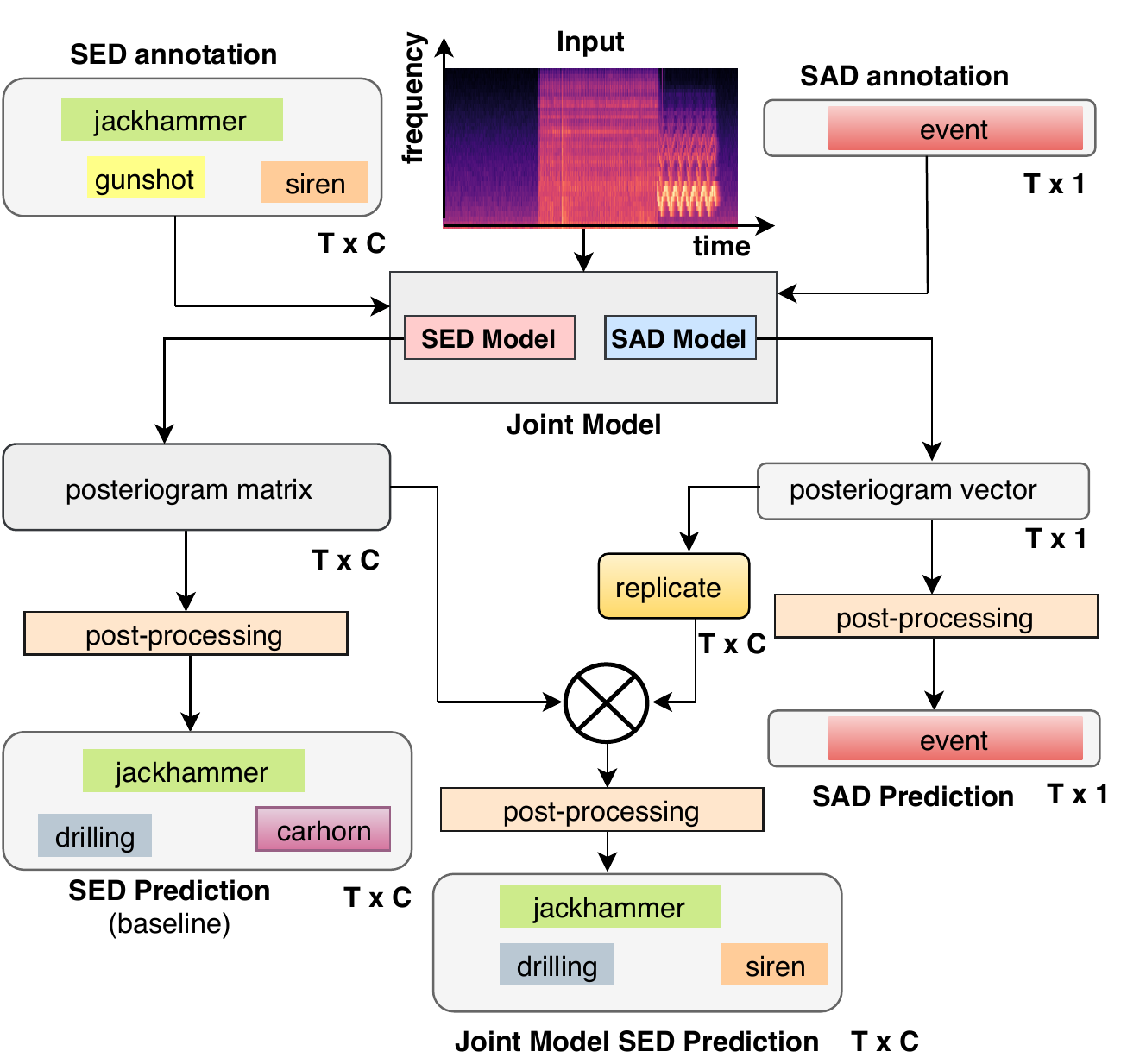}
  \caption{Block diagram of the proposed joint model for polyphonic SED ($T$ is the number of frames in the input data representation and $C$ is the total number of sound event classes in the dataset).}
  \label{fig:1}
\end{figure}

The majority of frame-wise prediction models proposed for polyphonic SED attempt to predict the class labels of sound events present in each frame of an audio sample. We assume that, to some extent, the majority of environmental sounds are either harmonic or percussive in nature. With this assumption, we propose that it is much easier to predict the presence or absence of any events in an audio frame, irrespective of categorization of events. In this work, we take advantage of the frame-wise training approach to propose a joint model for polyphonic SED using multi-task learning. The first task is polyphonic SED that predicts the class labels of sound events present in each frame of input audio. We call this task as `SED model', which is a frame based multi-label event classifier. The second task is an auxiliary model which is a binary classifier that predicts whether an event activity is present or not in each frame of the audio. We call this task as `SAD model'. From a signal processing perspective, we formulate the effects of dynamic polyphony levels in the sound events similar to the masking effects in speech. Consequently, low energy events are masked by high energy events at each frame. This results in many FN errors in event predictions which degrade the performance of SED. However SAD predictions are not affected by masking effects between co-occurring sound events. Furthermore, SAD can exploit polyphony to ensure the presence of an event even if one event is masked by another event with similar or different acoustic properties. The proposed joint approach takes advantage of the auxiliary branch to re-weight the prediction of sound events at each frame which helps to mitigate the FN errors in SED. A block diagram of the proposed joint model is shown in Fig.~\ref{fig:1}.

\subsection{Model configuration}
\label{ssec:subhead}

\begin{figure}[h!]
  \centering
  \centerline{\includegraphics[scale=0.6]{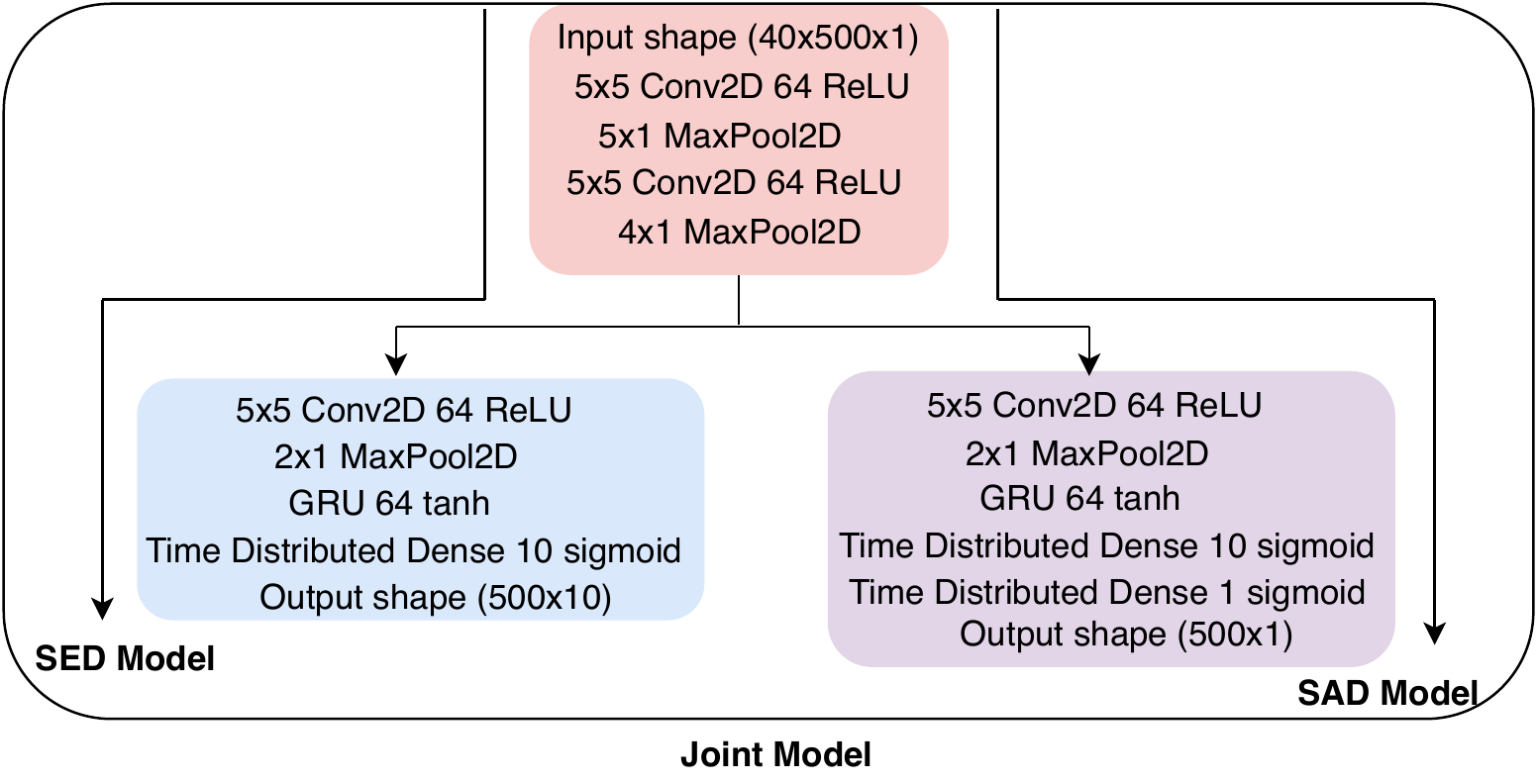}}
  \caption{The proposed joint model architecture along with the individual SED and SAD model architectures.}
  \label{fig:2}
\end{figure}

We use a state-of-the-art CRNN model architecture presented in \cite{cakir2017convolutional} to build our SED and SAD models, with some modifications. Our focus is to prove the effectiveness of the proposed joint model framework in enhancing polyphonic sound event detection. With this aim, we lessen the complexity of the CRNN architecture from the baseline architecture \cite{cakir2017convolutional}. Both the SED model and the SAD model have three blocks of convolutional layers, followed by a single Gated Recurrent Unit (GRU) layer. The SED model has a single dense layer after the GRU layer whereas the SAD model has two dense layers, which is the only difference between the two model architectures. The temporal dimension of the input data representation is unaltered in both the models. The detailed network architecture is shown in Fig.~\ref{fig:2}. The output of the SED model is a posteriogram matrix with dimensions $T\times C$, where $T$ is the number of frames in the input data representation and $C$ is the total number of sound event classes in the dataset. The output representation of the SAD model is a posteriogram vector with dimension $T$. The output of the joint model is the posteriogram matrix with dimensions $T\times C$, re-weighted on the SED output using the $T$ dimensional vector of the SAD model for each event class. This re-weighted posteriogram matrix is converted into a binary matrix using a threshold value prior to evaluation.

\subsection{Feature Extraction}
\label{ssec:subhead}

We use librosa \cite{mcfee2015librosa} to compute mel-scaled spectrograms from the input raw audio, which is the input data representation to all our models. Short-term Fourier transform (STFT) is employed to obtain the spectrogram from the input audio recordings with a hop length of $882$, an FFT window of $2048$, and a sample rate of 44.1kHz. This process converts a ten second (duration of recordings in the URBAN-SED \cite{salamon2017scaper} dataset) audio recording into a $1024 \times 500$ dimensional spectrogram representation. Each frame of this spectrogram is converted into a 40-dimensional vector of log filter bank energies using a Mel filterbank. We apply min-max normalization on the mel band energies. Hence, each 10-second audio recording is represented by a $40 \times 500$ dimensional Mel-spectrogram.

\subsection{Training}
\label{ssec:subhead}

We train both the SED model and the SAD model in a supervised manner. The dimension of the labels for the SED model is $T \times C$, and for the SAD model, it is $T$. The total loss of the joint model is the weighted sum of two cross-entropy losses; one from the SED model and one from the SAD model:
\begin{equation}
  \label{eqn:loss_equation}
    L_{joint\ model} = a\cdot L_{SED} + b\cdot L_{SAD}, 
\end{equation}
where, $L_{joint\ model}$ denotes the loss function of the joint model, and $L_{SED}$ and $L_{SAD}$ denote the cross entropy losses of the SED  and SAD models respectively. During training of the joint model, the individual losses are weighted differently to understand the influence of each task on the other; $a$ and $b$ denotes the respective loss weights. We also evaluate the influence of the number of shared layers on the joint model performance.

Batch normalization \cite{ioffe2015batch} is performed on the activations of every CNN layer and dropout \cite{srivastava2014dropout} with a probability of 0.30 is used for regularization. We train the network for 200 epochs using a binary cross entropy loss function for both tasks and with Adam \cite{kingma2014adam} optimizer with a learning rate of $0.001$. Early stopping is used to reduce the overfitting of the network to the training data. The proposed joint model is implemented using Keras \cite{keras} with Tensorflow \cite{abadi2016tensorflow} as backend.

\subsection{Experimental Setup}
\label{ssec:subhead}

To validate the effectiveness of the proposed joint model in improving polyphonic SED performance we conduct a total of four experiments. In all experiments, we train the neural network models with the same input data representation as described in subsection 2.2. The posteriogram outputs from the models are thresholded in order to obtain the binary event matrices prior to evaluation. We investigate the effect of different threshold values ($0.2$, $0.3$, $0.4$, $0.5$) on the performance of the standalone SED and SAD, and the joint model using the validation set. Based on the best results, we choose a threshold value of $0.2$ for the SED models in Exp. 1, 3, and 4. For the SAD model in Exp. 2 we choose $0.5$ as the threshold. The variables in capital bold letters refer to matrices and in small bold letters refer to vectors.

\textbf{Experiment 1}: We train a standalone SED model to evaluate SED performance. This is our baseline result. The model is trained with strong labels; $\textbf{GT}_{SED}$ with dimension $T$ $\times$ $C$ denotes the ground truth binary event activity matrix for each sample. Model architecture and training details are explained in subsection 2.1 and 2.3. $L_{SED}$ denotes the cross entropy loss of this model. For each sample, $\textbf{P}_{SED}$ with dimensions $T \times C$ is the posteriogram matrix output of the SED model. $\textbf{B}_{SED}$ denotes the binary equivalent of $\textbf{P}_{SED}$ after thresholding over a constant. Eventually, we use the binary event prediction matrices for evaluation.

\textbf{Experiment 2}: We train a standalone SAD model for event activity prediction as explained in subsection 2.3. The ground truth labels $\textbf{gt}_{SAD}$ used to train this model are the binary event activity vectors over the entire time span of the input data. The ground truth labels are created by truncating the strong label SED ground truths across the event classes as:
\begin{equation}
  \label{eqn:wave_equation}
    \textbf{gt}_{SAD} := \textbf{gt}_{{SED}_1}\lor
    \textbf{gt}_{{SED}_{2}}\lor \ldots \lor \textbf{gt}_{{SED}_{C}}
\end{equation}
where $\textbf{gt}_{{SED}_{C}}$ denotes the ground truth event activity vector corresponding to event $C$ of $\textbf{GT}_{SED}$. $L_{SAD}$ denotes the cross entropy loss of this model. For each sample, $\textbf{p}_{SAD}$ with dimension $T$ is the posteriogram vector output of the SAD model. $\textbf{b}_{SAD}$ denotes the binary equivalent of $\textbf{p}_{SAD}$ after thresholding over a constant. We use the binary event activity vectors for evaluation.


\textbf{Experiment 3}: To demonstrate the gain of carrying out sound activity detection on polyphonic SED performance, irrespective of the advantages of multi-task learning, we aggregate the output predictions of the standalone SED model of Exp. 1 and the SAD model of Exp. 2. The output posteriogram vectors from the SAD model are repeated to form posteriogram matrices as:
\begin{equation}
  \label{eqn:wave_equation}
    \textbf{P}_{SAD} := \text repeat \textbf{p}_{SAD}
\end{equation}
We use these posteriogram matrices to re-weight (Hadamard product) the outputs from the SED model to form the final predictions as:
\begin{equation}
  \label{eqn:wave_equation}
    \textbf{P}_{joint\_model} := \textbf{P}_{SED}\odot\textbf{P}_{SAD}
\end{equation}
The final predictions are thresholded to get event activity binary matrices to compute the evaluation metrics.

\textbf{Experiment 4}: The proposed joint model approach for polyphonic SED as explained in this section is trained in a multi-task learning setup. The total loss (eq. \ref{eqn:loss_equation}) of the model is the weighted sum of the SED model loss and the SAD model loss. Exp. 4a and 4b respectively represent the proposed joint model (with equal loss weights of $a$ = $b$ = $0.5$ and with loss weights of $a$ = $0.3$ and $b$ = $0.7$) for the SED and SAD tasks. Post-processing and evaluation is carried out as explained in Exp. 3.

\section{DATASET AND METRICS}
\label{sec:PDF_express}

We use the URBAN-SED \cite{salamon2017scaper} dataset in all experiments. URBAN-SED is a dataset of 10,000 soundscapes with sound event annotations generated using Scaper \cite{salamon2017scaper}, an open-source library for soundscape synthesis and augmentation. All recordings are of 10sec long, 16-bit mono and sampled at 44.1kHz. The annotations are strong, meaning for every sound event the annotations include the onset, offset, and label of the sound event. Every soundscape contains between 1-9 sound events from the list (air$\_$conditioner, car$\_$horn, children$\_$playing, dog$\_$bark, drilling, engine$\_$idling, gun$\_$shot, jackhammer, siren and street$\_$music) and has a background of Brownian noise. 
The URBAN-SED \cite{salamon2017scaper} dataset comes with pre-sorted train, validation and test sets; we use this default data split. Among 10,000 soundscapes, 6000 samples are used for training, 2000 samples for validation and 2000 samples for testing. 

In all experiments, we use the F-score and Error Rate (ER) for performance evaluation, with F-score as the primary evaluation metric. The evaluation metrics are computed in both segment-wise and event-wise manners using the sed\_eval tool \cite{mesaros2016metrics}. Segment-wise metrics compare system output and reference in short time segments. Event-wise metrics compare system output and corresponding reference event by event, with the metric showing how well the system detects event instances with the correct onset and offset. 
The evaluation scores presented in this work are micro averaged values, computed by aggregating intermediate statistics over all test data; each instance has equal influence on the final metric value. The ideal value of F-score is 100 and ER is zero. 
We use a segment length of one second to compute segment based metrics. The event-based metrics are calculated with respect to event instances by evaluating only onsets with a time collar of 250 ms.

\section{Evaluation}
\label{sec:copyright}


Both segment-based and event-based F-score and error rates are computed on the test set of the URBAN-SED \cite{salamon2017scaper} dataset for each of the experiments listed in Section 2. Tables \ref{table:1} and \ref{table:2} show the results of Exp. 4, which respectively summarize the influence of shared layers and loss weights on the performance of the joint model. To understand the effect of shared layers on joint model performance, the experiments are carried out with equal loss weights ($a$ = $b$ = $0.5$) as described in eq. (\ref{eqn:loss_equation}). The best scores (F-scores of $41.03\%$ (segment) and $8.76\%$ (event)) are achieved with sharing the first two convolutional layers. We use this model architecture to evaluate the influence of individual loss weights and in all further comparisons. Analysing the results of Table \ref{table:2}, the best metric scores are obtained by using the same weight of ($a$ = $b$ = $0.5$) for both the SED and the SAD losses. An interesting observation is that the event based metrics improve (from $8.76\%$ to $10.28\%$) when a slightly higher weight is given to the SAD loss which intuitively validates the influence of SAD task on efficient temporal localization of sound events.

Table \ref{table:3} summarizes the evaluation results for Experiments 1-4. Exp. 1 represents the baseline results of the standalone SED model. Exp. 2 shows the metric values of the standalone SAD model. The high scores of the SAD model (F-scores of $97.48\%$ and $43.14\%$ at segment and event levels) for event activity prediction justify the intuition of using the SAD model as an auxiliary branch to improve polyphonic SED performance. Exp. 3 represents the proposed joint model approach with separate training for the SED and SAD tasks. Analyzing the overall results, the segment based F-score increases from the baseline score of $35.48\%$ to $39.25\%$ and $41.03\%$ respectively for separate training and joint training of the proposed method. Similarly, there is improvement in event based F-score from a baseline value of $7.34\%$ to $11.13\%$ and $8.76\%$ respectively for separate training and joint training of the proposed method. With the proposed method error rates are also reduced at both segment and event levels. The event based F-score for the jointly trained model is slightly lower than the separately trained and aggregated model. However the improved F-score for segments and events for the separately trained model in Exp. 3 validates our method. With further improvements to the architecture and multi-task learning setup we believe the joint model can further improve detection performance. The results clearly indicate the effectiveness of the proposed method for improving polyphonic sound event detection performance.

\begin{table}[ht]
\caption{Results wrt number of shared layers for the joint model of Exp. 4.
\label{table:1}}
\centering 
\begin{tabular}{c c c c c} 
\hline 
& F1 (\%) &  & Error rate \\  
\hline 
Layers shared & Segment & Event & Segment & Event \\  
\hline 
 Conv 1 & 39.40 & 8 & 1.08 & 3.79 \\ 
 Conv 1,2 & \textbf{41.03} & \textbf{8.76} & \textbf{0.97} & \textbf{3.58} \\
 Conv 1,2,3 & 36.47 & 6.19 & 1.31 & 3.75 \\ 
\hline 
\end{tabular}
\label{table:nonlin} 
\end{table}

\begin{table}[ht]
\caption{Results wrt different loss weights for the model of Exp. 4.
\label{table:2}}
\centering 
\begin{tabular}{c c c c c} 
\hline 
loss weights & F1 (\%) &  & Error rate \\  
\hline 
(a, b) & Segment & Event & Segment & Event \\
\hline 
 (0.7, 0.3) & 36.12 & 5.62 & 1.08 & 4.41 \\ 
 (0.5, 0.5) & \textbf{41.03} & 8.76 & 0.97 & 3.58 \\
 (0.3, 0.7) & 34.18 & \textbf{10.28} & \textbf{0.91} & \textbf{2.36} \\ 
\hline 
\end{tabular}
\label{table:nonlin} 
\end{table}

\begin{table}[ht]
\caption{Results for Experiments 1, 2, 3, and 4.
\label{table:3}}
\centering 
\begin{tabular}{c c c c c} 
\hline 
& F1 (\%) &  & Error rate \\  
\hline 
Case & Segment & Event & Segment & Event \\ 
\hline 
 Exp. 1 & 35.48 & 7.34 & 1.54 & 3.81 \\ 
 Exp. 2 & 97.48 & 43.14 & 0.05 & 0.78 \\
 Exp. 3 & 39.25 & \textbf{11.13} & 1.21 & 2.90 \\
 Exp. 4a & \textbf{41.03} & 8.76 & 0.97 & 3.58 \\
 Exp. 4b & 34.18 & 10.28 & \textbf{0.91} & \textbf{2.36} \\
\hline 
\end{tabular}
\label{table:nonlin} 
\end{table}

Table \ref{table:4} shows another interesting observation from the best joint model of Exp. 4a. $J_{SAD}$ represents the sound activity detection performance of the `SAD model' of the joint model, similarly $J_{SED}$ and $J_{SED\_SAD}$ respectively represents sound event detection performance with the `SED model' and aggregation of predictions from `SED model' and `SAD model' of the joint model. The results show that with the best combination of loss weights and number of shared layers, the `SED model' of the joint model achieves almost the same results to that of the aggregation of the SED and SAD predictions. This means that with a good joint training procedure it is possible to achieve the best performance without the aggregation of individual SED and SAD predictions at a post-processing stage.

We also verify that the proposed method can lessen FP errors in both segments and events, and FN errors in events. We compute the segment-based and event-based precision~($P$) and recall~($R$) for Experiments 1, 3, and 4a (see Table~\ref{table:5}). Based on these results we also claim that to some extent the proposed approach helps in temporal localization of sound events. At the same time, we understand that our proposed method has the following limitations: 1) The proposed method is not very successful in decreasing the gap between segment-wise and event-wise scores in polyphonic SED; 2) Our work has not addressed the explicit modeling of sound events.

\begin{table}[ht]
\caption{Results on individual branches for the model of Exp. 4a.
\label{table:4}}
\centering 
\begin{tabular}{c c c c c} 
\hline 
& F1 (\%) &  & Error rate \\  
\hline 
Model & Segment & Event & Segment & Event \\ 
\hline 
 $J_{SAD}$ & 98.53 & 46.23 & 0.03 & 0.72 \\ 
 $J_{SED}$ & 40.99 & 8.28 & 0.97 & 3.65 \\
 $J_{SED\_SAD}$ & 41.03 & 8.76 & 0.97 & 3.58 \\
\hline 
\end{tabular}
\label{table:nonlin} 
\end{table}

\begin{table}[H]
\caption{Precision and Recall for SED for Exps. 1, 3, and 4a.
\label{table:5}}
\centering 
\begin{tabular}{c c c c c} 
\hline 
& P (\%) &  & R (\%) \\  
\hline 
Case & Segment & Event & Segment & Event \\  
\hline 
 Exp. 1 & 26.69 & 4.70 & \textbf{52.88} & 16.85 \\ 
 Exp. 3 & 31.82 & \textbf{7.57} & 51.19 & \textbf{21} \\
 Exp. 4a & \textbf{36.84} & 5.67 & 46.29 & 19.26 \\  
\hline 
\end{tabular}
\label{table:nonlin} 
\end{table}

\section{CONCLUSIONS AND FUTURE WORK}
\label{sec:pagelimit}
Within the limitations of current frame-based CRNN training methods, we proposed an auxiliary learning branch for event activity detection in order to improve polyphonic SED performance. We successfully evaluated the effectiveness of the proposed method on the URBAN-SED dataset. From our experimental results we conclude that: 1) the proposed joint model can improve polyphonic SED performance at both the segment and event levels and, 2) the proposed joint model can alleviate FP errors in both segments and events, and FN errors in events; which in turn improve the temporal localization of sound events. To further validate the method, we need to conduct similar experiments on more unbalanced and real-world datasets. We also plan to add one more branch to the existing method to predict frames conditioned on the event onsets as demonstrated in \cite{hawthorne2017onsets}. We hope such an implementation can bring down the error rate and also help in explicit modeling of sound events.

\end{document}